\documentstyle[12pt]{article}
\newcounter{appendixc} \renewcommand{\appendix}[1] {\vspace*{0.6cm}
        \refstepcounter{appendixc}
        \setcounter{figure}{0}
        \setcounter{table}{0}
        \setcounter{equation}{0}
        \renewcommand{\thefigure}{\Alph{appendixc}.\arabic{figure}}
        \renewcommand{\thetable}{\Alph{appendixc}.\arabic{table}}
        \renewcommand{\theappendixc}{\Alph{appendixc}}
        \renewcommand{\theequation}{\Alph{appendixc}.\arabic{equation}}
        \noindent{\bf Appendix #1}\par\vspace*{0.4cm}}
\begin{document}
\begin{titlepage} \begin{flushright} BIHEP-TH-98-10\\ \vspace{.3cm}
 \end{flushright} \vspace{0.1in} 
\begin{center} {\Large
Bounds On Anomalous Magnetic And Electric Moments }\end{center}
\begin{center} {\Large~ Of Tau Lepton From LEP And Lepton Flavor
Violation }
\vspace{.4in}

       Tao Huang, Zhi-Hai Lin, and Xinmin Zhang

\vspace{.4in} \it
 CCAST(World Laboratory) P.O. Box 8730, Beijing 100080, and\\
 Institute of High Energy Physics, Academia Sinica, P.O.Box 918\\
   Beijing 100039, China
\rm
\end{center}
\vspace{3cm}  

\begin{center} ABSTRACT\end{center}

The most stringent bounds on the anomalous magnetic and electric dipole
moments of the tau lepton was derived by
Escribano and Mass\'{o}[6] from the Z
width at LEP in an effective lagrangian approach to the new physics. In
this paper we point out that the higher dimensional
 operators introduced by
Escribano and Mass\'{o} not only modify the neutral currents of the tau lepton
to Z gauge boson and the photon, but also induce lepton flavor violation
(LFV). The size of the LFV effect depends crucially on the dynamics of the
lepton mass generation. Assuming the lepton mass matrices in the form of
an
Fritzsch ansatz, we point out that the experimental limit on $\mu
\rightarrow e \gamma$ will push the anomalous magnetic and electric
dipole moments of the tau lepton down to $10^{-11}$ and $10^{-25}~{\it e}
~{\rm cm}$ respectively.

\vfill PACS: 13.10.+q, 12.60.Cn

\end{titlepage} \eject
\baselineskip=0.30in 

Although the Standard Model (SM) has been successful in describing the
physics of the electroweak interaction[1], it is quite possible that the
SM is
only an effective theory which breaks down at higher energies as the
structure of the underlying physics emerges. There are reasons to
believe that the deviation from the SM might first appear in the
interactions involving the third-generation fermions[2]. In the lepton
sector, the tau lepton is
special in probing new physics since
it is the only lepton which is heavy enough to have hadronic
decays and it is generally believed that the heavier fermions are more
sensitive to the new physics
related to mass generation than light fermions. There
has been extensive
studies on the tau physics in the literature[3].  In particular, 
there has been growing interest in the electromagnetic and electric dipole
moments of the tau lepton[4][5] in recent years.
In general, the anomalous magnetic and the electric
dipole moments of the tau lepton are defined by (which we follow the
notation
of Ref.[6])
\begin{eqnarray}\label{eq1}
a_\tau = \frac{g_\tau -2 }{2} = F_2(q^2=0), ~~{\rm and} ~~d_\tau = e {\tilde
F}_2(q^2 = 0),
\end{eqnarray}
where 
$F_2$ and $\tilde F_2$ are form factors in the electromagnetic matrix element
\begin{eqnarray}\label{eq2}
< p^\prime | J_{em}^\mu(0) | p> = e {\bar u}(p^{\prime}) (F_1 \gamma^\mu +
                               [ \frac{i}{2 m_\tau} F_2 + \gamma_5 {\tilde F}_2]
                               \sigma^{\mu\nu} q_\nu ) u(p) ,
\end{eqnarray}
where $q = p^\prime - p $ and $ F_1(q^2=0 ) = 1$.

Theoretically the standard model predicts 
$a_\tau = 1.1769(4) \times 10^{-3}$ and a very tiny $d_\tau$[4] from CP
violation in the quark sector. Experimentally
recent analysis of the $e^+ e^- \rightarrow \tau^+ \tau^- \gamma$ process
from L3
and OPAL collaborations give that $a_\tau = 0.004 \pm 0.027 \pm 0.023$ and
$d_\tau = ( 0.0 \pm 1.5 \pm 1.3) \times 10^{-16} {\it e}~{\rm cm}$[5]. However
 the most
stringent bounds are that inferred from the width $\Gamma ( Z \rightarrow
\tau^+
\tau^- )$ [6], which are $-0.004 \leq a_\tau \leq 0.006$ and 
$| d_\tau | \leq 1.1 \times 10^{-17} {\it e}~{\rm cm}$.
To obtain these limits, Escribano and Mass\'{o} 
took an effective lagrangian approach to new physics,
\begin{eqnarray}\label{eq3}
{\cal L}_{eff} = {\cal L}_0 + \frac{1}{\Lambda^2} \sum_i C_i {\cal O}_i ,
\end{eqnarray}
where
${\cal L}_0$ is the SM Lagrangian, $\Lambda$ is the new physics
scale and $O_i$ are 
$SU_c(3)\times SU_L(2)\times U_Y(1)$ invariant operators. $C_i$ are
constants which represent the coupling strengths of $O_i$.
A complete list of CP-violating and CP-conserving operators have been given in
Ref.[7]. Regarding the anomalous magnetic moment of 
the tau lepton, the authors of
Ref.[6] have examined two dimension-six operators
\begin{eqnarray}\label{eq4}
{\cal O}_{\tau B}&=&{\bar L}\sigma^{\mu\nu} \tau_R \Phi B_{\mu\nu}, \\
{\cal O}_{\tau W}&=&{\bar L} \sigma^{\mu\nu} {\vec \sigma} \tau_R \Phi {\vec
W}_{\mu\nu},
\end{eqnarray}
where $L=( \nu_{\tau}, ~ \tau_L )$, $\Phi$ is the Higgs scalar, 
$B_{\mu\nu}$ and $W_{\mu\nu}$ are field strengths of $U_Y(1)$ and $SU_L(2)$.

When $\Phi$ gets vacuum expectation value,
operators ${\cal O}_{\tau B}$ and
${\cal O}_{\tau W}$ give rise to anomalous magnetic moment of the tau
lepton and also corrections to the decay width of $Z$ into ${\bar \tau}
\tau$. Given that the experimental data on Z width is quite consistent
with the
prediction of the SM, Escribano and Mass\'{o} put a strong bound on anomalous
magnetic moment of the tau lepton listed above. In this paper, we extend
the work of Ref.[6] by allowing the mixing of the three generations and
point out that the experimental limits on LFV will put stronger bounds than
Ref.[6] on anomalous magnetic and electric dipole moments of the tau
lepton. 

We present our arguments first with anomalous magnetic moment of the tau
lepton. Considering the dimension-six operators,  ${\cal O}_{\tau
B}$ and
${\cal O}_{\tau W}$, the full effective lagrangian, ${\cal L}_{eff}$ now
can be written as:
\begin{eqnarray}\label{eq5}
{\cal L}_{eff} = {\cal L}_0 + \frac{1}{\Lambda^2} ( c_{\tau B} {\cal O}_{\tau B}
    + c_{\tau W} {\cal O}_{\tau W} + h.c. ).
\end{eqnarray}
After the electroweak symmetry is broken and the mass matrices of the fermions and the
gauge bosons are diagonalized,
 the  
effective neutral current couplings of the leptons to gauge boson Z
 and the photon $\gamma$ are
\begin{eqnarray}\label{eq7}
{\cal L}^{Z, \gamma}_{eff}&=& eg^{Z, \gamma}
                    {{ \pmatrix{{\overline e}
  \cr {\overline \mu} \cr
  {\overline  
   \tau} \cr}}}^T \left \{
 ( \gamma_\mu V^{Z, \gamma} - \gamma_\mu \gamma_5 A^{Z, \gamma} )
-\frac{1}{ 2 m_\tau}( i k_\nu \sigma^{\mu\nu}) S^{Z, \gamma}
U_l
 \pmatrix{ 0 & & \cr
  & 0 & \cr
  & &  1 \cr }
  U_l^\dagger\right \}
  \pmatrix{e \cr \mu \cr \tau \cr}, \nonumber \\
& &
\end{eqnarray}
where $g^Z =  1/(4s_Wc_W), g^\gamma=1$, and
$V^{Z,\gamma}=1-4s_W^2,1$,
$A^{Z,\gamma}=1,0$ for Z and photon respectively, and 
\begin{eqnarray}
\label{eq32}
 S^Z&=&-\frac{8s_Wc_W}{e}\frac{m_\tau}{\Lambda^2}\frac{v}{\sqrt 2}\left [
   C_{\tau W }c_W-2C_{\tau B}s_W\right ],\\
\label{eq15}
 S^{\gamma}&=&\frac{2m_\tau}{e}\frac{\sqrt 2 v}{\Lambda^2}\left [
   C_{\tau W }\frac{s_W}{2}-C_{\tau B }c_W \right ].
\end{eqnarray}

The matrix $U_l$ in Eq.(7) is the unitary matrix which
diagonalizes the
mass matrix of the charged lepton.
In the SM, which corresponds to ${\cal L}_{eff}$ in the limit of $\Lambda
\rightarrow \infty$, the matrix $U_l$ is not
measurable because of the zero neutrino masses and furthermore,
 the universality of the
gauge interaction guarantees the absence
of the flavor changing
neutral current, the lepton flavor violation in the lepton sector.

The relative size of the $Z(\gamma) \tau {\bar \tau}$ to the flavor changing couplings
$Z (\gamma ) \tau {\bar \mu}$, {\it etc.}, in Eq.(\ref{eq7}) depends on the rotation matrix
$U_l$.
The elements of
$U_l$ can be evaluated
once the corresponding mass matrix is given. 
An interesting ansatz in the literature is the one 
 suggested by
Fritzsch and its variations [8-12]. The latter in Ref.[9] is given by
\begin{eqnarray}
U_l \; =\; \left ( \begin{array}{ccc}
c_{12}c_{13}                            & s_{12}c_{13}  
& s_{13} \\
-s_{12}c_{23}-c_{12}s_{23}s_{13}        & c_{12}c_{23}-s_{12}s_{23}s_{13}
& s_{23}c_{13} \\
s_{12}s_{23}-c_{12}c_{23}s_{13}         & -c_{12}s_{23}-s_{12}c_{23}s_{13}      
&c_{23}c_{13}
\end{array} \right ) ,\
\end{eqnarray}
where $s_{ij}\equiv \sin\theta_{ij}$, $c_{ij}\equiv \cos\theta_{ij}$ and
the three mixing angles are determined by corresponding lepton masses,
\begin{eqnarray}
\tan\theta_{12}&=&-1+\frac{2}{\sqrt{3}}\sqrt{\frac{m_{e}}{m_{\mu}}} ,\\
\tan\theta_{23}&=&-\sqrt{2}-\frac{3}{\sqrt{2}}\frac{m_{\mu}}{m_{\tau}},\\
\tan\theta_{13}&=&-\frac{2}{\sqrt{6}}\sqrt{\frac{m_{e}}{m_{\mu}}}.
\end{eqnarray}
With $U_l$ in Eqs.(10-13), we have
\begin{eqnarray}\label{eqvvv}
V_l = U_l \left (
\begin{array}{lcr}
0 &  & \\
& 0 & \\
 &  & 1
\end{array} 
\right )U_l^{\dagger}
&=&\left (
\begin{array}{lcr}
s_{13}^2           & s_{13}s_{23}c_{13} & s_{13}c_{23}c_{13} \\
s_{13}s_{23}c_{13} & s_{23}^2c_{13}^2 & s_{23}c_{13}^2c_{23} \\
s_{13}c_{23}c_{13} & s_{23}c_{13}^2c_{23} & c_{23}^2c_{13}^2 
\end{array}
\right ) \nonumber \\
&=&\left (
\begin{array}{lcr}
0.0035 & -0.049 &0.032 \\
-0.049 & 0.701  &-0.455 \\
0.032  & -0.455 &0.296 
\end{array}
\right ). 
\end{eqnarray} 
In the numerical calculation, we take that
 $\sqrt{m_{e}/m_{\mu}}\approx
0.0696$, and
$m_{\mu}/m_{\tau}\approx 0.0594$.
 
The decay width of $l\rightarrow l'+\gamma$ is
given by
\begin{eqnarray}\label{eq500}
\Gamma_{(l\rightarrow l'\gamma)}=
\frac{m_l}{32\pi}\left(V_{ll'}e S^{\gamma} \frac{m_l}{m_{\tau}}\right)^2,
\end{eqnarray}
where $V_{ll^{\prime}}$ ($l\not=l^{\prime}$) are the nondiagonal
elements of matrix $V$ defined in Eq. (14). In Eq. (15), We have neglected
the mass of the light lepton $l'$.

Given the current experimental upper limits on 
$\mu^- \rightarrow e^-\gamma$, $4.9
\times 10^{-11}$ [13], we have
\begin{eqnarray}
\label{eq55}
\vert S^{\gamma} \vert < 1.3 \times 10^{-10}. 
\end{eqnarray}

From Eqs.(7-9), the new physics contribution to the anomalous magnetic
moments of the tau lepton is given
by
\begin{eqnarray}\label{eq17}
\vert \delta \alpha_\tau \vert =
\left \vert V_{\tau \tau} S^{\gamma} \right \vert.
\end{eqnarray}
With the bounds on $S^\gamma$ in (16) and $V_{\tau \tau}$ in (14),
 we obtain that 
\begin{eqnarray}
\vert \delta \alpha_{\tau} \vert & \leq & 3.9 \times 10^{-11}.
\end{eqnarray}
This limit is much
stronger than that in Ref.[6]. The limits from other LFV processes,
such as $\tau^- \to e^- \gamma$, are weaker than that in Eq. (18).

Similarly, considering operators below which are introduced in Ref.[6],
\begin{eqnarray}
\label{eq23}
\tilde{\cal O}_{\tau B}&=&{\bar L}\sigma^{\mu\nu}i\gamma_5 \tau_R \Phi B_{\mu\nu}, \\
\label{24}
\tilde{\cal O}_{\tau W}&=&{\bar L}\sigma^{\mu\nu}i\gamma_5{\vec \sigma} \tau_R \Phi 
                   {\vec W}_{\mu\nu},
\end{eqnarray}
and following the procedure above in obtaining the bounds on tau lepton magnetic
moment, we put limit on the anomalous electric dipole moment 
\begin{eqnarray}
\vert d_{\tau} \vert & \leq & 2.2 \times 10^{-25}~e~{\rm cm}.
\end{eqnarray}
Again this is stronger than that obtained by Escribano and Mass\'{o}.

%
%
%


In summary, we extend the work by Escribano and Mass\'{o} to bound the anomalous magnetic
and
electric dipole moments of the tau lepton in the effective lagrangian by allowing the
mixing of three generations in the lepton sector. In
 the standard model, these mixing effects
are not measurable because of
vanishing neutrino masses and the universal
gauge interactions. With
non-universal interaction, the lepton flavor violation
 happens even with zero neutrino
masses. By taking the lepton mass matrix of Fritzsch ansatz,
 we have demonstrated that the
experimental limit on $\mu \rightarrow e \gamma$ put stronger
 limits on the anomalous
magnetic and electric dipole moments of the tau lepton than obtained
 by Escribano and Mass\'{o}
which is listed in the particle data book[13]. However 
our results depend on the
lepton mass matrix. So 
future experimental data on
anomalous magnetic and electric dipole moments of the tau lepton together with
lepton flavor violation will provide an experimental test on various
lepton mass ansatz.


\begin{center} {\Large Acknowledgments}\end{center}

We thank  Z. Z. Xing, J. M. Yang and B.-L. Young for
helpful discussions. This
work was supported in part by National Natural Science Foundation of
China.

\vspace{1cm} 
{\LARGE References}
\vspace{0.3in}
\begin{itemize}

\item[{\rm [1]}] R.D. Peccei, hep-ph/9811309.

\item[{\rm [2]}] R.D. Peccei and X. Zhang, Nucl Phys. {\bf B337}, 269
(1990); R. Peccei, S. Peris and X. Zhang, Nucl. Phys. {\rm B349}, 305
(1991); X. Zhang and B.-L. Young, Phys. Rev. {\bf D51}, 6564 (1995); H.
Georgi, L. Kaplan, D. Morin and A. Shenk, Phys. Rev. {\rm D51}, 3888
(1995); T. Han, K. Whisnant, B.-L. Young and X. Zhang, Phys. Lett. {\rm
B385}, 311 (1996); D. Carlson, E. Malkawi and C.-P. Yuan, Phys. Lett. {\rm
B337}, 145 (1994), G. Gounaris, M. Kuroda and F. Renard, Phys. Rev. 
{\rm55}, 5786 (12997).

\item[{\rm [3]}] For a review, see, A. Pich, hep-ph/9704453.

\item[{\rm [4]}] A. Czarnecki and W. Marciano, hep-ph/9810512.

\item[{\rm [5]}] L. Taylor,
hep-ph/9810463.

\item[{\rm [6]}] R. Escribano and E. Mass\'{o}, Phys. Lett. {\rm B395}, 369
(1997).

\item[{\rm [7]}] T. Huang, J.M. Yang, B.-L. Young and X. Zhang, Phys. Rev.
{\bf D58}, 073007 (1998); T. Huang, Z.-H. Lin and X. Zhang, in
preparation.

\item[{\rm [8]}]
                H. Fritzsch, Nucl. Phys. B {\bf 155} (1979) 189.
\item[{\rm [9]}] H. Fritzsch and Z. Z. Xing, 
                 Phys. Lett. B {\bf  372} (1996) 265.
\item[{\rm [10]}]H. Fritzsch and J. Plankl,
                 Phys. Lett. B {\bf 237} (1990) 451;
                 H. Fritzsch and D .Holtmannsp\"{o}tter,
                 Phys. Lett. B {\bf 338} (1994) 290;
                 H. Fritzsch and Z. Z. Xing, hep-ph/9808272,  
                 to be published in Phys. Lett. B;
                 H. Fritzsch and Z. Z. Xing, hep-ph/9807234;
                 M. Fukugita, M. Tanimoto and T. Yanagida,
                 Phys. Rev. D {\bf 57}, (1998) 4429;
                 M. Tanimoto, hep-ph/9807283.
\item[{\rm [11]}] K. Kang, S. K. Kang, C. S. Kim and S. M. Kim, 
		  hep-ph/9808419 
\item[{\rm [12]}] Z. Xing, hep-ph/9804433.
\item[{\rm [13]}] C.Caso et al., Particle Data Group, Eur. Phys. J. 
		  {\bf C 3} (1998) 1.
\end{itemize}
\eject

\end{document}